\documentclass[graybox]{svmult}

\usepackage{mathptmx}       % selects Times Roman as basic font
\usepackage{helvet}         % selects Helvetica as sans-serif font
\usepackage{courier}        % selects Courier as typewriter font
\usepackage{type1cm}        % activate if the above 3 fonts are
\usepackage{makeidx}         % allows index generation
\usepackage{graphicx}        % standard LaTeX graphics tool
\usepackage{multicol}        % used for the two-column index
\usepackage[bottom]{footmisc}% places footnotes at page bottom

\newcommand{\simgreat}{\mathbin{\lower 3pt\hbox
   {$\rlap{\raise 5pt\hbox{$\char'076$}}\mathchar"7218$}}}
\newcommand{\simless}{\mathbin{\lower 3pt\hbox
   {$\rlap{\raise 5pt\hbox{$\char'074$}}\mathchar"7218$}}}

\bibliography{referenc.tex}

\makeindex             % used for the subject index
\begin{document}

\title*{Guilt by Association: Finding Cosmic Ray Sources Using Hierarchical Bayesian Clustering}
\titlerunning{ Finding Cosmic Ray Sources}
\author{Kunlaya Soiaporn\and David Chernoff \and Thomas Loredo \and David Ruppert \and Ira Wasserman}
\institute{Kunlaya Soiaporn \at Cornell Unversity, Ithaca,NY, \email{ks354@cornelll.edu}
\and David Chernoff \at Cornell Unversity, Ithaca,NY \email{chernoff@astro.cornell.edu}}
\authorrunning{K. Soiaporn et al.}
\maketitle

\abstract*{The Earth is continuously showered by charged cosmic ray particles, naturally 
produced atomic nuclei moving with velocity close to the speed of light.
Among these are ultra high energy cosmic ray particles with energy 
exceeding $5\times 10^{19}$ eV, which is ten million times more energetic than the most 
energetic particles produced at the Large Hadron Collider.
Astrophysical questions include: what phenomenon accelerates particles to such high energies, and 
what sort of nuclei are energized? Also, the magnetic deflection of the trajectories of 
the cosmic rays makes them potential probes of galactic and intergalactic magnetic fields.
We develop a Bayesian hierarchical model that can be used to compare different 
association models between the cosmic rays and source population, using Bayes factors. 
A measurement model with directional uncertainties and accounting for non-uniform 
sky exposure is incoporated into the model. The methodology allows us to learn 
about astrophysical parameters, such as those governing the source luminosity function and the cosmic magnetic field.}

\section{Introduction}
\label{sec:1}
Since the Pierre Auger Observatory (PAO) initiated observations in 2004 it
has detected
14 ultra high energy cosmic rays (UHECRs) with energy $\geq 55$ Eev
in period 1--January 1, 2004 - May 26, 2006, 13 UHECRs in  
period 2-- May 27, 2006 - August 31, 2007, and
42 UHECRs in period 3-- September 1, 2007 - December 31, 2009.
The energy threshold of 55 Eev was chosen by using period 1 data \cite{PAO}.
These CR particles interact with the cosmic microwave background, and 
according to GZK limit, CRs with energy $\simgreat$ 60 Eev should come from
sources within 200 Mpc \cite{PAO}. We consider the 17 active galactic nuclei (AGNs)
in the volume-complete (to 15 Mpc) catalog of \cite{AGN} as candidate sources. 
We use a Bayesian hierarchical 
model to compare 3 models, M$_0$: only isotropic background source, 
M$_1$: isotropic background + 17 AGNs, M$_2$: isotropic background 
+ 2 AGNs (Centaurus A and NGC 4945--the two 
closest AGNs) for the UHECRs from the three periods.

\section{Models and Algorithms}
\label{sec:2}
We describe the CR arrival as a Poisson process with rate
set by source fluxes and exposure factors,
the measurement error
as a Fisher distribution with the angular uncertainty 
of 0.9$^\circ$ and the magnetic deflection as a Fisher distribution with 
concentration parameter $\kappa$.
Our hierarchical model has parameters $F_0$ (flux from isotropic background), 
$F_A$ (total flux from the AGNs), $\lambda$ (source label of each UHECR), 
and $\kappa$. We assume AGNs have fixed CR luminosity implying an AGN
at distance $d$ generates CR flux $\propto 1/d^2$.
We analyze a physically plausible range of deflections
$\kappa\in [1,1000]$. $F_0$ 
and $F_A$ have an exponential prior with scale 
$s \approx 0.063$ km$^{-2}$ yr$^{-1}$, based on previous data from CR
observatories AGASA and HiRes. Gibbs sampling is performed on the parameters 
$F_A, F_0$ and $\lambda$ to obtain the posterior distributions.
We use Chib's method in \cite{Chib} to estimate the marginal likelihood under each model.

\section{Results}
\label{sec:3}
\begin{figure}
\centerline{$
\begin{array}{cc}
\includegraphics[width=1.5in,angle=-90]{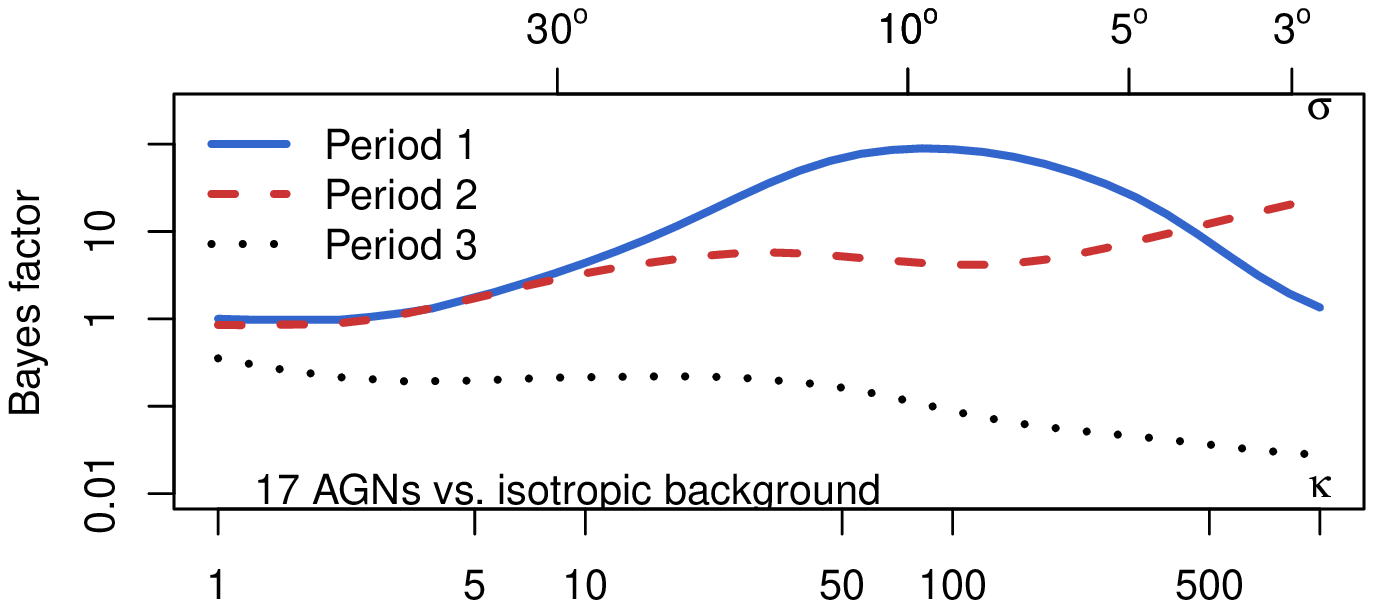} &
\includegraphics[width=1.5in,angle=-90]{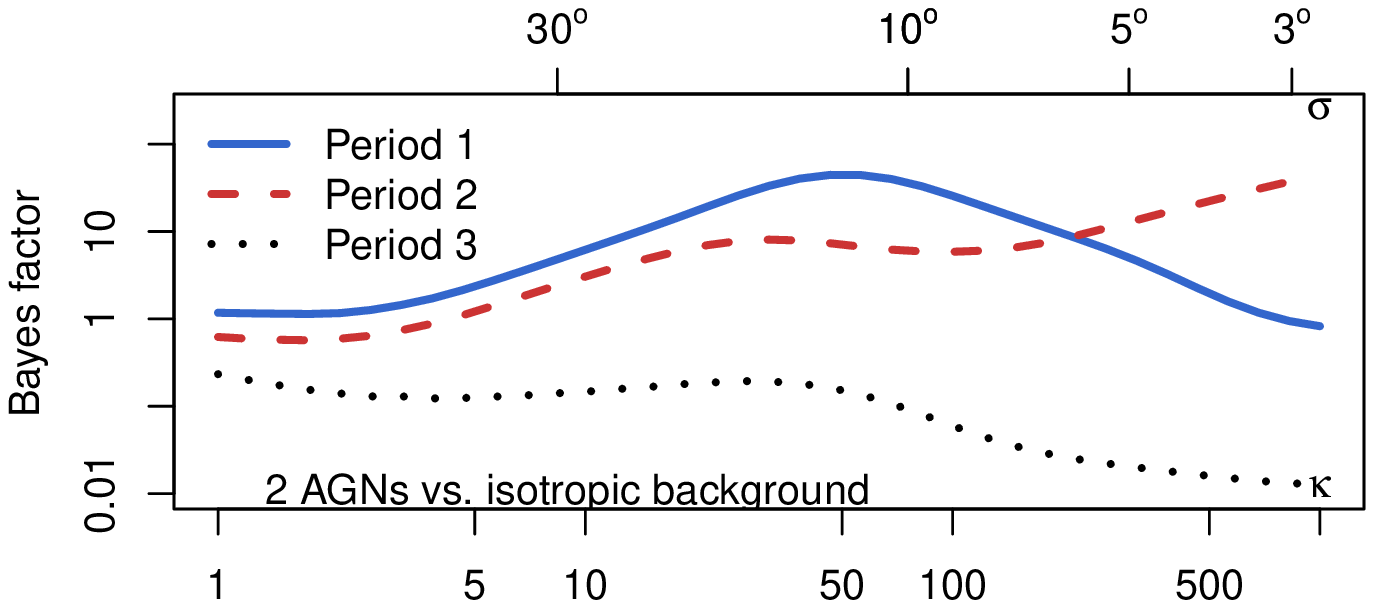}
\end{array}$}
\caption{Bayes factors comparing the association model with 17 AGNs (left) or 2 AGNs (right) to the null isotropic background model. 
$\sigma$ is the standard deviation in 2-d Gaussian approximation for the Fisher distribution }
\label{fig:bayesfactor}
\end{figure}

The Bayes factors as a function of $\kappa$ are shown in Fig.~\ref{fig:bayesfactor}. Adopting the log-flat prior for $\kappa$,
we obtain the overall Bayes factor against the null of $26.10$, $5.41$ and $0.15$ for M$_1$
and $12.37$, $8.27$ and $0.11$ for M$_2$, for periods 1, 2 and 3, respectively. The
strength of the evidence for AGN association differs markedly from period to period. For M$_1$ and M$_2$ we find $\simless$ 10\% of PAO CRs may come from AGN and a significant fraction must
originate elsewhere.
%%%%%%%%%%%%%%%%%%%%%%%% referenc.tex %%%%%%%%%%%%%%%%%%%%%%%%%%%%%%
% sample references
% %
% Use this file as a template for your own input.
%
%%%%%%%%%%%%%%%%%%%%%%%% Springer-Verlag %%%%%%%%%%%%%%%%%%%%%%%%%%
%
% BibTeX users please use
% \bibliographystyle{}
% \bibliography{}

\begin{thebibliography}{9.}%
% Use the following syntax and markup for your references if 
% the subject of your book is from the field 
% "Mathematics, Physics, Statistics, Computer Science"
%

\bibitem{PAO}The Pierre Auger Collaboration, Abreu, P., et al. (2010). Update on the Correlation of the Highest Energy Cosmic Rays with Nearby Extragalactic Matter. Astroparticle Physics, 34(5):314-.

\bibitem{Chib}Chib, S. (1995). Marginal Likelihood from the Gibbs Output. Journal of the American Statistical Association, 90(432):1313-1321.

\bibitem{AGN}Goulding, A.D., Alexander, D.M., Lehmer, B., Mullaney, J.R.(2010). Towards a Complete Census of Active Galactic Nuclei in Nearby Galaxies: the Incidence of Growing Black Holes. Monthly Notices of the Royal Astronomical Society, 406(1):597-61.

\end{thebibliography}
%

\end{document}